\documentclass[%
 reprint,
 amsmath,amssymb,
 aps,
 prl,
]{revtex4-1}

\usepackage{graphicx}
\usepackage{dcolumn}
\usepackage{bm}
\usepackage{float}
\usepackage{siunitx} 
\usepackage{xcolor}
\usepackage[colorlinks=false,hidelinks]{hyperref}

\begin{document}

\preprint{APS/123-QED}

\title{Controlled Injection in a Laser Plasma Accelerator via an Optically Generated Waveguide Constriction}

\author{R. J. Shalloo}
\email{rob.shalloo@desy.de}
\affiliation{Deutsches Elektronen-Synchrotron DESY, Notkestraße 85, 22607 Hamburg, Germany}%

\author{A. Ferran Pousa}
\affiliation{Deutsches Elektronen-Synchrotron DESY, Notkestraße 85, 22607 Hamburg, Germany}%

\author{M. Mewes}
\affiliation{Deutsches Elektronen-Synchrotron DESY, Notkestraße 85, 22607 Hamburg, Germany}%

\author{S. Jalas}
\affiliation{Deutsches Elektronen-Synchrotron DESY, Notkestraße 85, 22607 Hamburg, Germany}%

\author{M. Kirchen}
\affiliation{Deutsches Elektronen-Synchrotron DESY, Notkestraße 85, 22607 Hamburg, Germany}%

\author{R. D'Arcy}
\affiliation{Deutsches Elektronen-Synchrotron DESY, Notkestraße 85, 22607 Hamburg, Germany}%

\author{J. Osterhoff}
\affiliation{Deutsches Elektronen-Synchrotron DESY, Notkestraße 85, 22607 Hamburg, Germany}%

\author{K. P\~oder}
\affiliation{Deutsches Elektronen-Synchrotron DESY, Notkestraße 85, 22607 Hamburg, Germany}%

\author{M. Th\'evenet}
\affiliation{Deutsches Elektronen-Synchrotron DESY, Notkestraße 85, 22607 Hamburg, Germany}%

\date{\today}

\begin{abstract}
We propose a novel scheme for controlling the injection of a high-quality electron bunch into a channel-guided laser plasma accelerator. 
This all-optical technique, constricted waveguide injection, creates a highly tunable controlled injection structure natively within a plasma waveguide, a key requirement for efficient acceleration of high-quality multi-GeV electron beams. 
We describe a simple optical setup to tailor the plasma and present start-to-end simulations showing the injection structure formation and the generation of a \SI{1.1}{\giga\electronvolt} electron beam with \SI{10}{\pico\coulomb} of charge and 0.35 \% energy spread using \SI{1}{\joule} of drive laser energy. 
Highly tunable tailored plasma sources, like those proposed here, enable fine control over the injection and acceleration processes and thus will be crucial for the development of application-focused laser plasma accelerators.
\end{abstract}

\keywords{Laser Plasma Accelerator, Wakefield, HOFI, Controlled Injection}

\maketitle
Laser plasma accelerators (LPAs) \cite{Tajima1979} are leading a new generation of ultra-compact particle and photon sources owing to their ability to sustain accelerating fields on the scale of hundreds of \si{\giga\volt/\metre} --- three orders of magnitude higher than radio-frequency accelerators \cite{Hooker2013}. 
This dramatic reduction in machine size opens up the possibility of distributing compact GeV-level accelerators not just at large-scale national laboratories, but at universities, hospitals and factories, greatly increasing the availability of beams for applications in areas of high societal impact such as photon science, particle physics, medicine and industry.

Plasma density tailoring is crucial for improving LPA performance. 
Longitudinal (along the laser axis) density tailoring has enabled high-quality electron beam generation \cite{Geddes2008,Schmid2010,Gonsalves2011,vGrafenstein2023} suitable for driving a UV-FEL \cite{Wang2021} while transverse tailoring has been used to create laser guiding structures \cite{Durfee1993,Spence2003} leading to multi-GeV energy gains \cite{Leemans2006,Leemans2014,Gonsalves2019}.
More recently, an exciting push to combine longitudinal and transverse tailoring has led to multi-GeV electron beams with just a few percent energy spread \cite{Oubrerie2022,Picksley2023,Picksley2024}.
However, the generation of beams for applications such as LPA-driven x-ray FELs or synchrotron injector systems remains challenging. 
Significant developments in plasma source technology including advanced plasma density tailoring methods, injection concepts, and modelling processes will be critical to improving the quality of plasma-accelerated beams. 

In this Letter, we propose a method of controlling injection within a plasma waveguide to enable the generation of GeV-level electron beams with sub-percent energy spread and sub-micron emittance.
The all-optical technique, constricted waveguide injection (CWI), seeds injection by engineering a localized rapid variation in the waveguide structure. 
For the work presented here, the required 3D tailoring of the plasma density profile is facilitated through the hydrodynamic expansion of multiple bodies of optical-field-ionized plasma. 
The process is examined via start-to-end simulations exploring the formation of the plasma waveguide structure with a localized constriction, the controlled injection of electrons into the plasma wake, and the subsequent acceleration process.

The injection structure is formed as a modification to a hydrodynamic optical-field-ionized (HOFI) plasma waveguide \cite{Shalloo2018,Lemos2013,Lemos2013a}. In such a waveguide, a column of plasma is ionized and heated via optical field ionization. This hot column of plasma then expands radially outwards, driving a shock wave into the cold surrounding gas. The resulting plasma density profile has been shown to be suitable for guiding high-intensity laser pulses over meter-scale distances \cite{Shalloo2019,Picksley2020,Picksley2020a,Feder2020,Miao2020}, all with low axial densities ($\sim10^{17}$ \si{\per\centi\metre\cubed}) commensurate with multi-GeV electron energy gain \cite{Miao2022}.

\begin{figure} 
	\centering
	\includegraphics[width=\columnwidth]{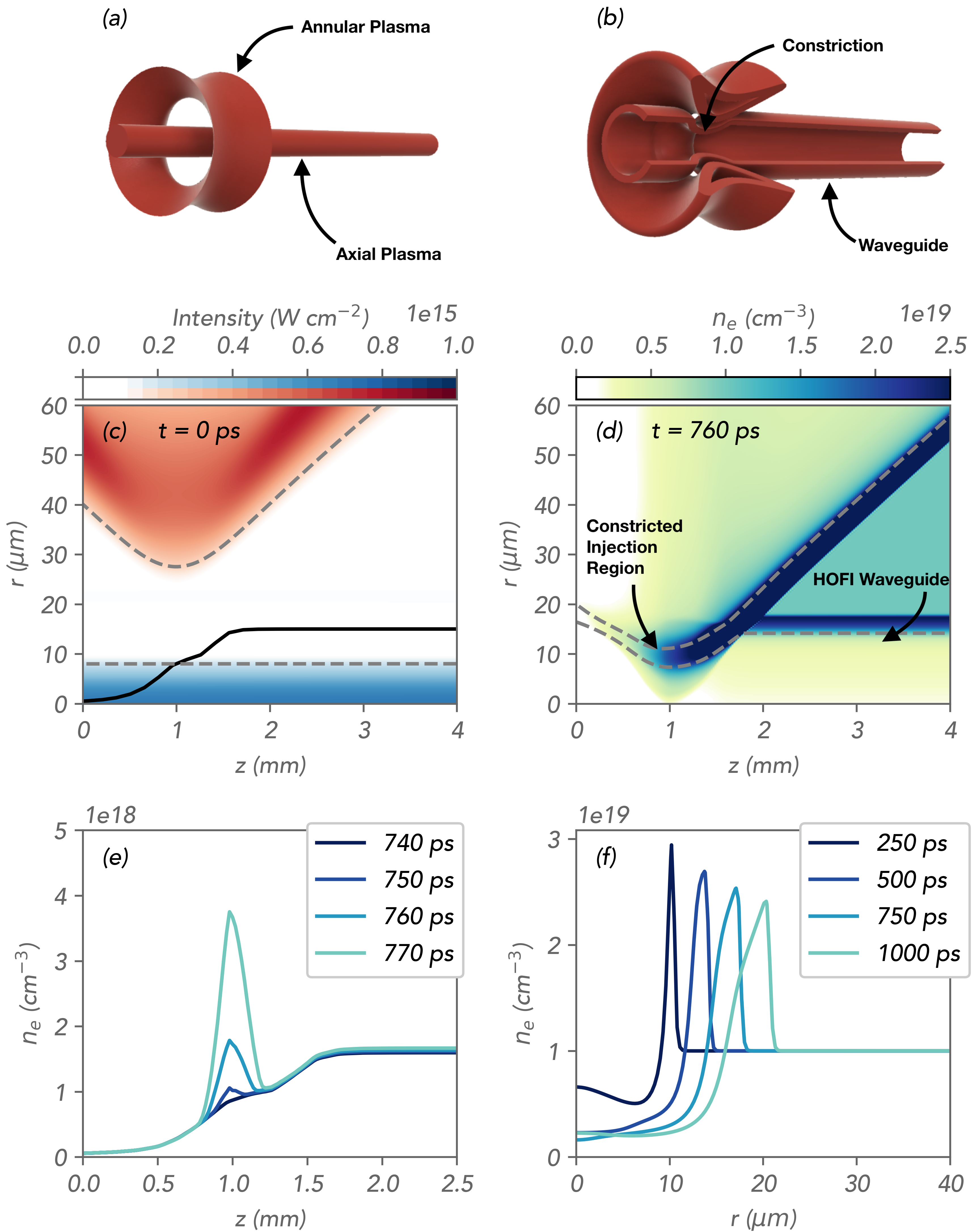}
	\caption{Concept: (a) initial plasma formed (b) plasma after expansion. Simulations of structure formation: (c) shows the r-z intensity profile created by a \SI{400}{\nano\metre} axicon beam (blue) and \SI{800}{\nano\metre} helical lens beam (red). The black line shows the neutral gas density profile. The dashed gray line denotes the boundary over which the ionisation fraction drops to 10 \%. (d) Total electron density (assuming ionization of remaining neutrals) after \SI{760}{\pico\second} of evolution. Plasma evolution: (e) axial electron density in steps of \SI{10}{\pico\second} and (f) radial electron density at $z=$ \SI{4}{\milli\meter} in steps of \SI{250}{\pico\second}.} \label{fig:concept}
\end{figure}

In addition to ionizing a plasma column for waveguide formation, here we form an additional annular plasma concentric with the plasma column and longitudinally located near the beginning of the accelerator. After ionization, these two plasma structures hydrodynamically expand and interact. 
The axial plasma column expands radially outwards to form a HOFI waveguide while the annular plasma both expands outwards and converges inwards where it acts to locally modify the structure of the waveguide. 

The structure formation is presented in Fig. \ref{fig:concept} with (a) and (b) elucidating the formation of the plasma structure conceptually.
Figure \ref{fig:concept} (a) illustrates the initial plasma and (b) represents the plasma after expansion with the waveguide being modified in the region of the initial plasma annulus.
The plasma source formation simulations follow the multiphysics simulation framework described in Ref. \cite{Mewes2023}, although with a 2D cylindrically symmetric geometry to account for longitudinal effects in the vicinity of the constriction.
The intensity of the two plasma-forming beams is shown in Fig. \ref{fig:concept} (c) together with the neutral hydrogen gas density profile assumed for the simulations (calculated from a gas cell CAD model using \textsc{ANSYS Fluent} \cite{ansys}). 
In the configuration shown here, an axicon is used to ionize a plasma column while a helical lens (combination of lens and azimuthal phase plate) is used to ionize an annular plasma. 
Figure \ref{fig:concept} (d) shows the electron density after almost a nanosecond of hydrodynamic evolution, assuming ionization of the remaining neutrals by the leading edge of a guided pulse \cite{Picksley2020a,Feder2020}. 
Far from the pinched region, a HOFI waveguide is formed, while in the vicinity of the annulus the waveguide structure has been modified, showing rapid longitudinal changes in the axial density, channel radius and channel shape. 
As will be shown, this modified region can be effective at promoting localized electron injection leading to the generation of high-quality electron beams. Additionally, as the structure is natively formed within a plasma waveguide, these high-quality beams can be subsequently accelerated to GeV energies.

Figure \ref{fig:concept} (e) and (f) show respectively the evolution of the axial density in the vicinity of the constriction and the radial density far from the constriction.  
While the plasma waveguide expands on nanosecond timescales, the converging annular plasma modifies the waveguide properties on picosecond timescales due to the accelerated flow velocity of the plasma as it approaches the optical axis. 
These plasma evolution dynamics are beneficial for LPAs as three critical processes are spread across three different timescales; plasma waveguide formation on nanosecond timescales, injection structure generation and tuning on picosecond timescales, and electron acceleration on femtosecond timescales. 

The plasma dynamics, and so the final 3D plasma structure, can be tuned by controlling the initial conditions of each body of plasma (temperature, density profile) through appropriate choice of laser parameters (wavelength, pulse duration, energy, polarization, etc.) and by adjusting the relative formation times of each body of plasma. Additionally, the use of tailored neutral gas density profiles and dopant species can be considered.

In the following, to further explore this technique, we provide detailed start-to-end simulations of the plasma source formation and subsequent laser plasma acceleration process. 
Simulations of the plasma source formation follow the methodology described above while the laser plasma acceleration process was simulated using a combination of two cylindrical codes: the particle-in-cell code \textsc{FBPIC}~\cite{Lehe2016,Kirchen2020} is used to simulate the beam injection, and the subsequent acceleration is modelled in the quasi-static code \textsc{Wake-T}~\cite{FerranPousa2019} to strongly reduce the simulation cost. 
The \emph{handshake} between the two codes is facilitated by the \textsc{LASY} library~\cite{Thevenet2024}, such that the results are in agreement with simulations performed with \textsc{FBPIC} alone. This pipeline enabled an optimization and tuning process via an extensive simulation campaign of thousands of simulations. The key simulation parameters --- drive laser focal position $z_f$ and spot size $w_0$, plasma length $l_p$, and plasma expansion time $\tau_p$--- were optimised using Bayesian Optimization with the \textsc{Optimas} library \cite{FerranPousa2023,Hudson2022}. Top-scoring working points indicate a trade-off between electron beam quality, energy and charge. We hereafter focus on a $\sim 1$~GeV case.

\begin{figure} 
    \centering
    \includegraphics[width=\columnwidth]{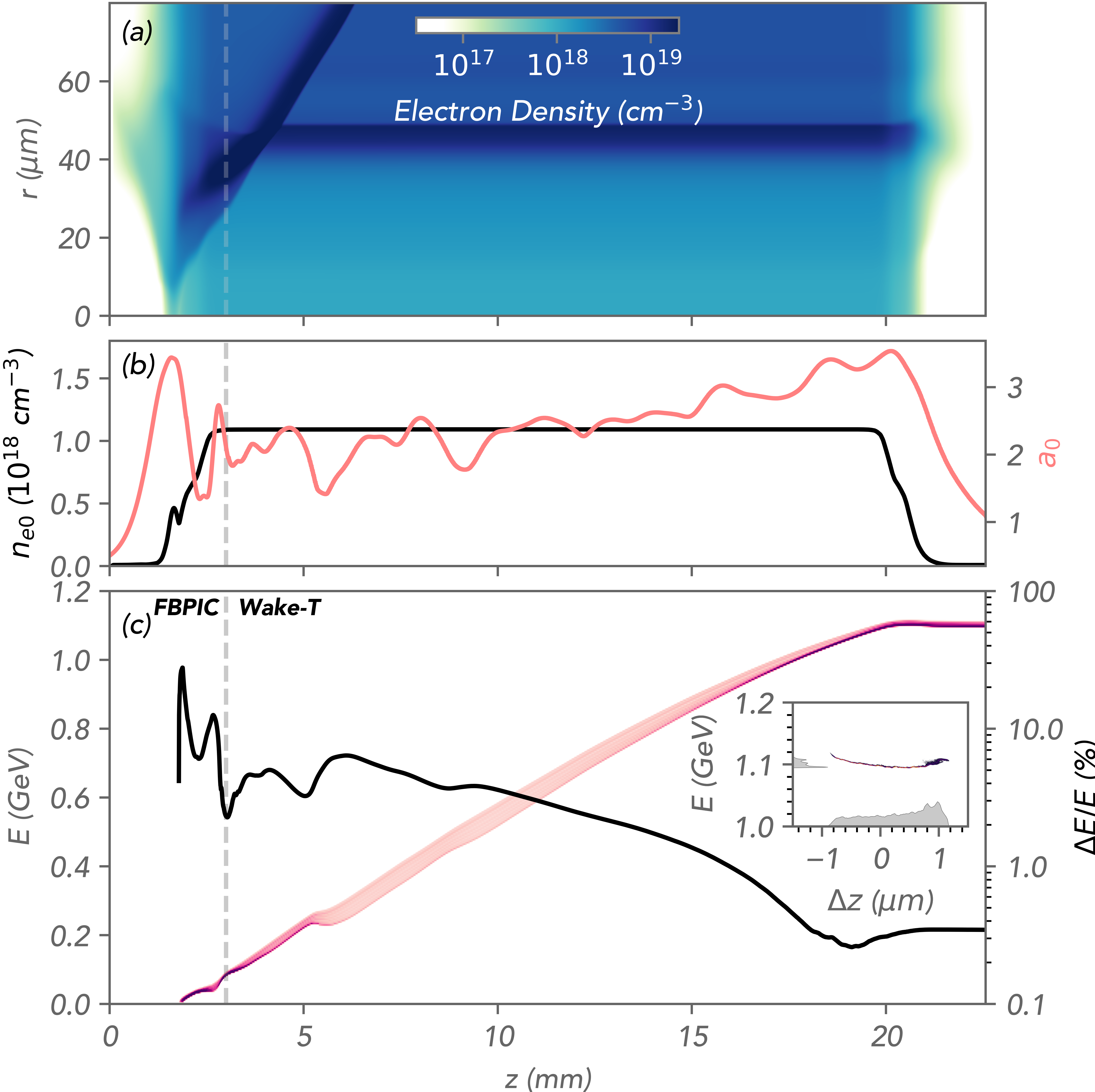}
    \caption{(a) Total electron density profile at \SI{1.156}{\nano\second}. (b) The on-axis electron density (black) together with the evolution of the laser strength (red). (c) Evolution of the electron energy spectrum (pink) and spread (black). The grey dashed line on each subplot represents the handshake between \textsc{FBPIC} and \textsc{Wake-T}.}\label{fig:densityAndEvolution}
\end{figure}

An \SI{800}{\nano\meter}, \SI{27}{\femto\second} laser pulse is used for simultaneous formation of both channel and injection structure. 
The channel is formed using a \SI{0.5}{\degree} approach angle axicon while the injection structure is created via an $m=16$ helical lens with an f-number of 25. 
The energy required for channel and injection structure formation is less than \SI{50}{\milli\joule} in these simulations. 
In a practical implementation the required energy is expected to be slightly higher due to effects such as ionization-induced refraction \cite{Leemans1992,Leemans1992a}, which is not modelled here. 
The plasma structures are formed in hydrogen with a uniform transverse profile and a varying longitudinal profile as described above. 
In the simulation coordinates, the helical lens focal position is located at $z=$ \SI{2}{\milli\meter}, within the rising density ramp of the gas cell profile. 
For the simulations presented here, a plasma expansion time $\tau_p = 1.156$  \si{\nano\second} was found to be optimal. 
This time coincided with the moment at which the collapsing annular plasma began to reach the optical axis. 
This generated a localized constriction in the plasma waveguide entrance at $z \sim$ \SI{1.65}{\milli\meter}, slightly upstream of the helical lens focus as the plasma expansion speed increases with a reduction in the ambient gas density.
The r-z density profile after expansion is shown in Fig. \ref{fig:densityAndEvolution} (a) together with an axial lineout in Fig. \ref{fig:densityAndEvolution} (b).

The acceleration process was driven by a \SI{1}{\joule}, \SI{800}{\nano\meter}, laser pulse which in the selected case was focused to a spot with a vacuum $1/e^2$ intensity radius of \SI{9.9}{\micro\meter} and a normalized laser vector potential $a_0=3.09$. The vacuum focus was located at a distance $z_f=$ \SI{1.5}{\milli\meter} along the propagation axis. 
On entering the plasma structure, the laser strength peaks at $a_0=3.44$ in the channel constriction. 
As the laser pulse leaves the constriction, there is a rapid expansion of the plasma bubble which facilitates injection. 
The peak vector potential of the laser throughout the accelerator is shown in Fig. \ref{fig:densityAndEvolution} (b) while the evolution of the electron beam energy spectrum is presented in Fig. \ref{fig:densityAndEvolution} (c).

The final electron bunch (inset Fig. \ref{fig:densityAndEvolution} (c)) comprises \SI{10}{\pico\coulomb} of charge with a mean energy of \SI{1.1}{\giga\electronvolt} and a mean absolute deviation (MAD) energy spread of \SI{3.8}{\mega\electronvolt}, leading to a 0.35\% energy spread. 
The slice-averaged energy spread is 0.05\%. 
As noted previously, the MAD has been found to be a robust statistical measure of energy spread for LPA electron beams \cite{Jalas2021}. 
For comparison, the FWHM and RMS energy spreads are 0.32\% and 0.46 \% respectively. 
The resulting beam has a normalized projected ($x \,, y$) emittance of $(0.37 \,, 0.12)$ \si{\micro\meter}.

An understanding of the injection process can be obtained by considering the bubble dynamics as the laser pulse leaves the constricted region of the plasma waveguide.
The speed of the front of the bubble is locked to the laser pulse, which travels at its group velocity in the plasma structure. 
Injection occurs due to a reduction in the speed of the back of the plasma bubble, which is caused by the elongation of the bubble \cite{Bulanov1998,Suk2001}. 
Typically, this slow-down is associated with a density down-ramp, or an increase in laser strength. 
However, in contrast to previous injection mechanisms, here the on-axis density $n_0$ increases overall and the normalized vector potential $a_0$ decreases throughout the injection process, thus both acting to speed up the back of the bubble and hinder injection. 
Instead, the overall deceleration of the back of the plasma bubble seen here occurs due to the rapid change in the transverse density profile of the plasma source and the expansion of the laser into this new structure.

\begin{figure}[!ht]
    \centering
    \includegraphics[width=\columnwidth]{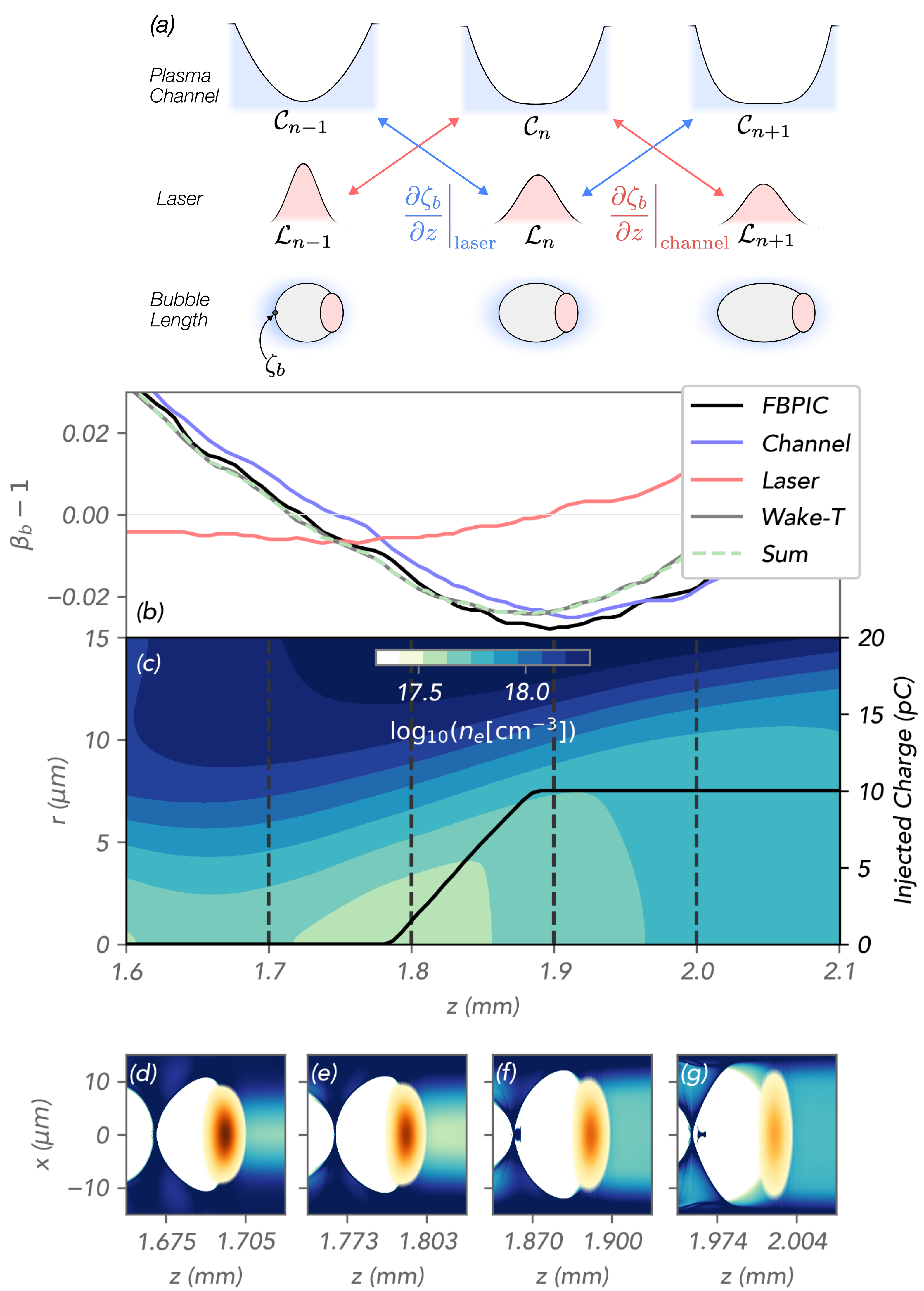}
    \caption{(a) Method to determine the respective contributions of the laser and channel evolution to the bubble elongation. (b) Speed of the back of the plasma bubble in \textsc{FBPIC} and \textsc{Wake-T} and the speed due to the laser and channel evolution individually, together with their sum. (c) Contour plot of the r-z density profile at the exit of the constriction. The solid line shows the injection of the accelerated beam. The dashed lines indicate the four locations at which (d)-(g) are plotted. Subplots (d)-(g) show the evolution of the plasma bubble and the injection of an electron beam.} \label{fig:injectionSimulation}
\end{figure}

To understand the individual contributions of both the laser and the plasma channel evolution to the injection process one can consider the speed of the back of the plasma bubble $\beta_b$, in the absence of injected charge. This can be expressed as the sum of the changes due to the channel (laser fixed) $\partial \zeta_b /\partial z |_{\mathrm{laser}}$ and due to the laser driver (channel fixed) $\partial \zeta_b / \partial z |_{\mathrm{channel}}$ separately:
\begin{equation}
\beta_b -1 = \frac{d \zeta_b}{dz} =\, \frac{\partial \zeta_b}{\partial z}\bigg |_{\mathrm{laser}} + \frac{\partial \zeta_b}{\partial z}\bigg |_{\mathrm{channel}},
\label{eqn:speed}
\end{equation}
where $\zeta_{b}$ is the location of the back of the bubble relative to the moving window (co-moving variable).

\textsc{Wake-T} can compute the plasma response for given laser and channel profiles, without tracking the plasma evolution self-consistently. We exploit this feature of quasi-static methods to evaluate numerically these partial derivatives by computing the plasma response, and so $\zeta_{b}$, for various combinations of laser and channel profiles. 
The method is illustrated in Fig. \ref{fig:injectionSimulation} (a) and can be expressed as
$\left (\partial \zeta_b / \partial z |_{\mathrm{laser}}\right)_n \simeq \left[ \zeta_{b} (\mathcal{L}_n,\mathcal{C}_{n+1} ) - \zeta_{b} (\mathcal{L}_n,\mathcal{C}_{n-1} )\right] /(2 \Delta z)$, where $\mathcal{L}_{n}$ is the laser profile extracted from \textsc{FBPIC} at iteration $n$, and $\mathcal{C}_{n}$ is the channel profile at the corresponding longitudinal position $z=nc\Delta t$ obtained from the hydrodynamic simulation. Similarly, $\left (\partial \zeta_b / \partial z |_{\mathrm{channel}} \right)_n \simeq \left[ \zeta_{b} (\mathcal{L}_{n+1},\mathcal{C}_{n} ) - \zeta_{b} (\mathcal{L}_{n-1},\mathcal{C}_{n} )\right]/(2 \Delta z)$. In practice, at each timestep during the injection process, the laser pulse is extracted from the \textsc{FBPIC} simulation and read by \textsc{Wake-T} using \textsc{LASY}, where it can be combined with an arbitrarily chosen channel profile to calculate the plasma response.

Figure \ref{fig:injectionSimulation} (b) tracks the speed of the back of the bubble throughout the injection process. The contribution of the plasma channel and the laser to the speed is shown together with their sum, which is in agreement with the speed of the back of the bubble measured from a single \textsc{Wake-T} simulation, indicating that Eq. \ref{eqn:speed} is numerically satisfied to good precision. These results also match the \textsc{FBPIC} simulation up until the point at which significant charge has been injected. 
Here, it is seen that the dominant factor in slowing down the back of the bubble is the change in the plasma channel profile.

These bubble dynamics result from the rapid variation in the transverse density profile as the plasma source transitions from constriction to uniform guide (Fig. \ref{fig:injectionSimulation} (c)).
As the laser leaves the constricted region the channel becomes larger with a steeper wall. 
Consequently, the density locally flattens in the vicinity of the laser pulse, decreasing the restoring force experienced by electrons in the sheath, leading to a bubble elongation and an initiation of the injection process.
It is observed that cessation of injection occurs due to changes in the bubble shape in response to the modal evolution of the laser as it expands to fill the larger uniform section of the waveguide. 
Figure \ref{fig:injectionSimulation} (d)-(g) show snapshots of the \textsc{FBPIC} simulation illustrating the bubble elongation and injection process.

For the example presented here, injection is facilitated through the rapid change in waveguide structure including on-axis density, shape and width. 
The injection method does not depend on obstructions in a high-velocity neutral gas flow, e.g. wire/blade and gas jet based schemes, which place a high load on vacuum pumps and can limit the repetition rate of such systems. Additionally, the scheme does not require femtosecond scale timing between laser pulses which can be practically challenging.
Finally the method creates a tunable injection structure natively within a HOFI waveguide and can be used to generate low emittance and low energy spread beams at GeV energies.
We note here that previous work in the context of capillary discharge waveguides has investigated a separate scheme in which varying only the waveguide matched spot size leads to off-axis injection of electron beams suitable for a tunable betratron x-ray source \cite{Liu2018}. 

In summary, we have presented a new concept for controlling the injection of electrons into the wake of a channel-guided LPA. 
The constricted waveguide injection method relies on engineering a rapid variation in waveguide properties to control the speed of the back of the plasma bubble. 
The 3D tailoring of the plasma source is achieved via hydrodynamic plasma structuring which offers a high degree of tunability.
Start-to-end simulations show an example case demonstrating the injection and subsequent acceleration of a \SI{10}{\pico\coulomb}, \SI{1.1}{\giga\electronvolt} electron beam with a mean absolute deviation energy spread of 0.35\% and sub-micron projected emittance using a \SI{1}{\joule} drive laser pulse.
The injection dynamics are investigated in detail and the contribution of both laser and plasma isolated. 
It is shown that the plasma channel evolution is the dominant factor driving the injection process. 
Here we have demonstrated that structuring the plasma to tailor the waveguide properties can enable new techniques and enhanced control over the plasma acceleration process. 
As plasma accelerator beam quality requirements grow more demanding, high-fidelity plasma structuring will become an increasingly important tool.
Looking to the future, the use of a collapsing annular shock, without waveguide structure, could find utility as an all-optical injection technique for beam-driven plasma accelerators.
In the context of LPAs, two pertinent examples where plasma structuring techniques could find application include the generation of multi-GeV electron beams for use as i) a compact x-ray FEL driver and ii) an injector for a conventional storage ring \cite{Antipov2021}.

\begin{acknowledgments}
This work was funded by the Deutsche Forschungsgemeinschaft (DFG, German Research Foundation) – Projektnummers 531352484 and 491245950.
This research was supported by the Maxwell computational resources operated at Deutsches Elektronen-Synchrotron
DESY, Hamburg, Germany.
The authors gratefully acknowledge the Gauss
Centre for Supercomputing e.V. (www.gauss-centre.eu) for funding this project by providing computing time through the John von Neumann Institute for Computing (NIC) on the GCS Supercomputer JUWELS at J\"{u}lich Supercomputing Centre (JSC).
The authors would like to thank Andreas Maier for useful discussions.

\end{acknowledgments}

%
    
\end{document}